\documentclass{piparticle-final}

\usepackage{graphicx}
\usepackage{amsmath}
\usepackage{amsfonts}
\usepackage{amssymb}
\usepackage{chemarrow}
\usepackage{chemarr}
\usepackage{wasysym}
\usepackage{dsfont}

\usepackage{cite} 
\usepackage{color} 
\usepackage{epstopdf} 

\definecolor{rmarker}{rgb}{0.9,0.0,0.0}
\definecolor{gmarker}{rgb}{0.0,0.7,0.0}
\definecolor{bmarker}{rgb}{0.1,0.1,0.9}
\definecolor{cmarker}{rgb}{0.0,0.7,0.7}

\begin{document}

\volume{6}               
\articlenumber{060012}   
\journalyear{2014}       
\editor{G. Mindlin}   
\received{3 August 2014}     
\accepted{28 October 2014}   
\runningauthor{I. M. Lengyel \itshape{et al.}}  
\doi{060012}         

\title{Nonlinearity arising from noncooperative transcription factor binding enhances negative feedback and promotes genetic oscillations}

\author{Iv\'an M. Lengyel,\cite{inst1}
	Daniele Soroldoni,\cite{inst2,inst3}
	Andrew C. Oates,\cite{inst2,inst3}
	Luis G. Morelli\cite{inst1}\thanks{Email: morelli@df.uba.ar}}

\pipabstract{
We study the effects of multiple binding sites in the promoter of a genetic oscillator. We evaluate the regulatory function of a promoter with multiple binding sites in the absence of cooperative binding, and consider different hypotheses for how the number of bound repressors affects transcription rate. Effective Hill exponents of the resulting regulatory functions reveal an increase in the nonlinearity of the feedback with the number of binding sites. We identify optimal configurations that maximize the nonlinearity of the feedback. We use a generic model of a biochemical oscillator to show that this increased nonlinearity is reflected in enhanced oscillations, with larger amplitudes over wider oscillatory ranges. Although the study is motivated by genetic oscillations in the zebrafish segmentation clock, our findings may reveal a general principle for gene regulation.
}

\maketitle

\blfootnote{
\begin{theaffiliation}{99}
	\institution{inst1} Departamento de F\'{\i}sica, FCEyN UBA and IFIBA, CONICET, Pabell\'on 1, Ciudad Universitaria, 1428 Buenos Aires, Argentina.
	\institution{inst2} MRC-National Institute for Medical Research, The Ridgeway, Mill Hill, NW7 1AA London, UK.
	\institution{inst3} Department of Cell and Developmental Biology, University College London, Gower Street, WC1E 6BT London, UK.
\end{theaffiliation}
}


\section{Introduction}

\begin{figure}[b]
\begin{center}
\includegraphics[width=\columnwidth]{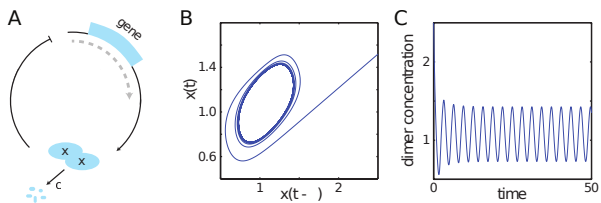}
\caption{Delayed autoinhibition can produce genetic oscillations. 
(A) The gene (light blue box) is transcribed and translated into gene products x, with a delay $\tau$. 
In this example, gene products form dimers that act as transcriptional repressors, inhibiting transcription of the gene (blunted arrow), and decay at a rate $c$.
(B) and (C) Numerical solutions of Eq.~(\ref{eq.oscillator}) describing the oscillator in A.
(B) Phase space: monomer concentration $x(t)$ vs. the delayed concentration $x(t-\tau)$ settled to a limit cycle.
(C) Dimer concentration oscillates as a function of time.
Parameters in B, C: $b\,P = 2$, $\tau=1$, $c=1$, $x_0=1$, $N=12$, $M=6$.}
\label{fig.loop}
\end{center}
\end{figure}

Cells can generate temporal patterns of activity by means of genetic oscillations~\cite{goldbeter,mackey}.
Genetic oscillations are biochemical oscillations in the levels of gene products~\cite{liu95,nagoshi04,mihalcescu04,goldbeter12,palmeirim97,aulehla08,masamizu06,krol11,shimojo08,geva06,elowitz00,stricker08}.
They can be produced by negative feedback regulation of gene expression, in which a gene product inhibits its own production directly or indirectly~\cite{tyson.in.keizer}.
Such autoinhibition is often performed by transcriptional repressors, proteins or protein complexes that bind the promoter of a gene and inhibit the transcription of new mRNA molecules~\cite{alberts,alon06}, see Fig.~\ref{fig.loop}.
A theoretical description of biochemical oscillations requires a delayed negative feedback, together with sufficient nonlinearity and a balance of the timescales of the different processes involved~\cite{novak08}.
Delays may occur naturally in transcriptional regulation, since the assembly of gene products involves intermediate steps~\cite{alberts,lewis03,morelli07}.
Nonlinearity refers to the presence of nonlinear terms in the equations describing the dynamics.
Such nonlinear terms may occur in the equations due to the presence of cooperative biochemical processes, 
where cooperativity is understood as a phenomenon in which several components act together to orchestrate some collective behavior~\cite{ferrell09}.
Although some processes giving rise to nonlinear terms are known, in general it is still an open question how nonlinearity is built into genetic oscillators. 

A compelling model system for genetic oscillations is the vertebrate segmentation clock~\cite{pourquie11,oates12,saga12b,roellig11}.
This is a tissue-level pattern generator that controls the formation of vertebrate segments during embryonic development~\cite{roellig11,saga12a}.
The spatiotemporal patterns generated by the segmentation clock are thought to be initiated by a genetic oscillator at the single cell level~\cite{palmeirim97,oates02,hirata02,holley02,masamizu06,krol11}.
In this genetic oscillator, negative feedback is provided by genes of the Her family, 
which code proteins that form dimers and can act as transcriptional repressors~\cite{takebayashi94,bessho01,schroter12,trofka12,hanisch13}. 
The time taken from transcription, translation and splicing may introduce the necessary feedback delays in zebrafish and mouse~\cite{lewis03,giudicelli07,ozbudak08,hanisch13,harima13}.
One source of nonlinearity in the segmentation clock oscillator is the dimerization of gene products that bind the promoter of cyclic genes~\cite{schroter12}.
However, this may be insufficient to generate the observed oscillations in the levels of gene products.
One way to increase nonlinearity would be cooperative binding of repressors to regulatory binding sites at the promoter~\cite{keener,ferrell09,qian12}. 
Cooperative binding to multiple binding sites can make the negative feedback steeper~\cite{zeiser06}.
A similar effect occurs with ultrasensitivity in phosphorylation cascades~\cite{gunawardena05}.
The presence of clusters of binding sites for transcription factors may be a common motif in gene regulation~\cite{gotea10},
and cooperativity in transcription factor binding has been reported for some systems~\cite{burz98}.

In zebrafish, multiple binding sites for Her dimers have been identified in the promoter region 
of Her1, Her7, and other genes of the Notch pathway~\cite{brend09,schroter12,soroldoniPC}. 
However, there is no evidence for cooperative binding of transcriptional regulators in the case of the zebrafish segmentation clock.
Therefore, although cooperative binding is not ruled out, this lack of evidence raises the general question of 
what contribution could be expected from the multiple binding sites reported in the promoter of segmentation clock cyclic genes.
Here we use theory to study how multiple binding sites affect nonlinearity and biochemical oscillations in a generic description of a genetic oscillator.

\section{A promoter with multiple binding sites}
\begin{figure}[b]
\begin{center}
\includegraphics[width=\columnwidth]{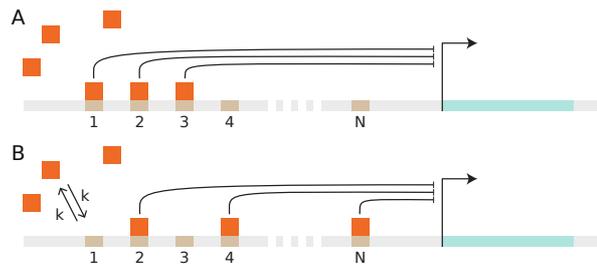}
\caption{Schematic representation of a promoter with multiple binding sites, numbered from $1$ to $N$.
Transcriptional repressors (orange squares) bind and unbind from binding sites (numbered platforms) at the promoter of a gene (light blue stretch). 
(A) and (B) illustrate two equivalent configurations of the promoter with identical number of bound transcriptional repressors having the same inhibiting strength.
}
\label{fig.promoter}
\end{center}
\end{figure}

We first evaluate the regulatory function of a promoter that contains $N$ binding sites for a transcriptional repressor, see Fig.~\ref{fig.promoter}.
We consider transcriptional repressors although the results in this section are more general and would apply to other types of transcription factors.
We focus on a single transcriptional repressor for the sake of simplicity, and assume that all binding sites are identical.
That is, binding and unbinding of repressors to the different sites occur at the same rates for all sites.
Moreover, we assume that there is no cooperativity in repressor binding.
This means that binding and unbinding rates are not affected by the presence of bound factors to any of the other sites. 

The state ${\cal P}_{i}$ of the promoter at any time can be characterized by the number $i$ of bound factors, which goes from $0$ to $N$.
With a rate $k^{+}$, a free repressor binds to an empty site at the promoter, which steps from state ${\cal P}_{i}$ to state ${\cal P}_{i+1}$;
with a rate $k^{-}$, a bound repressor falls off the promoter, which steps from state ${\cal P}_{i}$ to state ${\cal P}_{i-1}$:

\begin{align}
{\cal P}_{0} \xrightleftharpoons[k^{-}]{k^{+}} &\cdots \xrightleftharpoons[]{} {\cal P}_{i-1} \xrightleftharpoons[k^{-}]{k^{+}} {\cal P}_{i} \xrightleftharpoons[k^{-}]{k^{+}} {\cal P}_{i+1} \xrightleftharpoons[]{} \notag \\
&\cdots \xrightleftharpoons[k^{-}]{k^{+}} {\cal P}_{N}.
\end{align}
Denoting by $P_i$ the promoter occupation probabilities which describe the fraction of time that the promoter spends at state ${\cal P}_i$ in the thermodynamic limit~\cite{bintu05a},
the kinetics of binding and unbinding of transcriptional repressors to the binding sites is given by the set of ordinary differential equations

\begin{eqnarray}
\dot P_{0} &=& - k^{+} N x P_{0} + k^{-} P_{1}					\nonumber	\\
&\vdots&												\nonumber	\\
\dot P_{i}	&=& - k^{+} (N-i) x P_{i} - k^{-} i P_{i} 				\\
{}		&{ }& + k^{+} (N-i+1) x P_{i-1} + k^{-} (i+1) P_{i+1}			\nonumber\label{eq.odes}	\\
&\vdots&												\nonumber	\\
\dot P_{N} &=& - k^{-} N P_{N} + k^{+} x P_{N-1}	\, ,			\nonumber
\end{eqnarray}
together with the conservation law

\begin{equation}
P = \sum_{i=0}^{N}{P_{i}}	\, ,							\label{eq.conservation}
\end{equation}
where $x$ is the repressor concentration and $P$ is proportional to the number of gene copies.

We assume here that binding and unbinding of repressors to the promoter occur much faster than other processes like transcription, translation, transport and decay of molecules.
This means that the promoter occupation probabilities quickly reach equilibrium with a given concentration of transcriptional repressors~\cite{bintu05a,garcia10}.
This situation is described by $\dot P_{i} = 0$ for all $i$ in Eq.~(\ref{eq.odes}), and the resulting algebraic equations can be solved by induction to obtain

\begin{equation}
P_{i} = {N \choose i} \left( \frac{x}{x_0} \right)^{i} P_{0}	\, ,	\label{eq.pipo}
\end{equation}
where $x_0 = k^{-} / k^{+}$ is the equilibrium constant for binding of factors.
Using the constraint Eq.~(\ref{eq.conservation}) we express $P_{0}$ in terms of $P$

\begin{equation}
P_{0} = P \left( 1 + \frac{x}{x_0} \right)^{-N}			\, .	\label{eq.pop}
\end{equation}
Equation~(\ref{eq.pipo}) and Eq.~(\ref{eq.pop}) describe the equilibrium occupation of the promoter in terms of the concentration $x$ of the transcriptional repressor.

\section{Abrupt inhibition}
\begin{figure}[t]
\begin{center}
\includegraphics[width=\columnwidth]{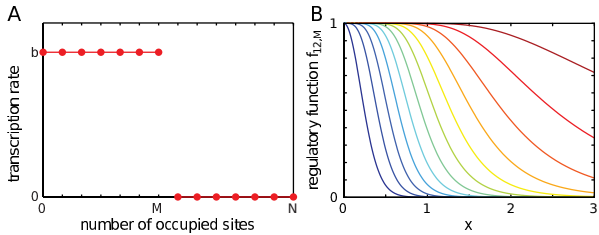}
\caption{Abrupt inhibition. Full inhibition occurs when more than $M$ sites are bound by transcriptional repressors.
(A) Transcription rate as a function of the number of bound transcriptional repressors. 
(B) Normalized regulatory function, Eq.~(\ref{eq.fnm}), as a function of repressor concentration $x$, with $N=12$ and $M = 0,\ldots,11$  (dark blue to dark red).
}
\label{fig.step}
\end{center}
\end{figure}

In the previous section, we evaluated the kinetics of noncooperative binding to a promoter with multiple binding sites.
How does the presence of bound transcriptional repressors affect the transcription rate of the gene downstream of the promoter?
In general, the strength of inhibition will depend on the number of bound repressors, and the regulatory function $f(x)$ will have the form

\begin{equation}		\label{eq.fx}
f(x) = \sum_{i=0}^{N}{a_i P_{i}} \, ,
\end{equation}
where $a_0=b$ is the basal transcription rate in the absence of bound repressors and $a_i$ is the transcription rate in the presence of $i$ bound repressors.
Here we assume, for the sake of simplicity, that transcription proceeds at its basal rate $b$ while there are $M$ or less sites occupied by the repressors, 
and drops to zero when the number of occupied sites is larger than $M$, see Fig.~\ref{fig.step}. We shall consider an alternative scenario below.

In this situation, Eq.~(\ref{eq.fx}) becomes

\begin{equation}		\label{eq.fnm0}
f(x) = b \sum_{i=0}^{M}{P_{i}} \, .
\end{equation}

Using Eq.~(\ref{eq.pipo}) and Eq.~(\ref{eq.pop}), the resulting regulatory function for a promoter with $N$ binding sites is

\begin{align}		\label{eq.fnm}
f_{N,M}(x) &= b P \left( 1 + \frac{x}{x_0} \right)^{-N} \notag\\
&\times \sum_{i=0}^{M}{  {N \choose i} \left( \frac{x}{x_0} \right)^{i}  } \, .
\end{align}

\begin{figure}[t]
\begin{center}
\includegraphics[width=\columnwidth]{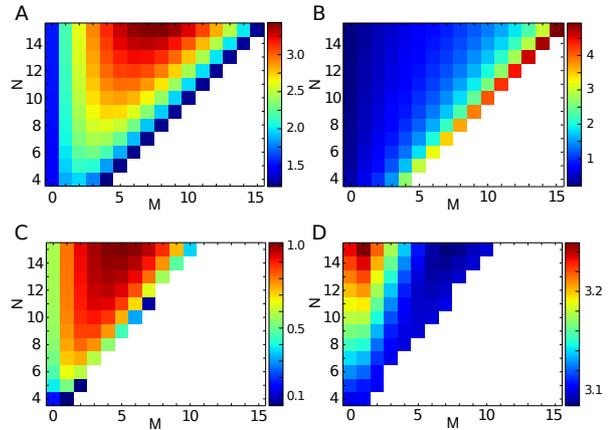}
\caption{Multiple binding sites can increase nonlinearity and enhance oscillations.
(A) and (B) Effective Hill parameters for the regulatory function $f_{N,M}$, Eq.~(\ref{eq.fnm}), with $b\,P=1$ and $x_0=1$.
(A) Effective Hill coefficient $h$.
(B) Effective inhibition threshold $K$.
(C) and (D) Oscillations described by Eq.~(\ref{eq.oscillator}) with $b\,P= 2$, $\tau=1$, $c=1$ and $x_0=1$.
(C) Amplitude of oscillations.
(D) Period of oscillations.
In C and D, the white region represents the nonoscillatory regime, in which the system settles to a fixed point.
Color bar labels indicate values in each panel.
}
\label{fig.step_fit}
\end{center}
\end{figure}

The zebrafish segmentation clock may be an interesting system to evaluate these results.
In the \emph{her1/her7} locus of zebrafish, an estimated number of about $12$ binding sites has been reported~\cite{schroter12,soroldoniPC}.
The regulatory functions resulting from Eq.~(\ref{eq.fnm}) for $N = 12$ are displayed in Fig.~\ref{fig.step}B.
Although it is clear that inhibition of the promoter for a given level of repressors shifts to the right as $M$ increases,
it is less obvious how the value of $M$ affects the steepness ---that is the nonlinearity--- of the negative feedback.
We use Hill functions to parametrize the regulatory functions Eq.~(\ref{eq.fnm}) in a more transparent form, see Appendix. 
Hill functions are parametrized by a Hill coefficient $h$ characterizing the steepness of the curve, and an inhibition threshold $K$ that describes the concentration of repressors that halves the production rate.
We fit Hill functions to $f_{N,M}$ and obtain an effective Hill coefficient $h$ and effective inhibition threshold $K$, for each value of $N$ and $M$, Fig.~\ref{fig.step_fit}A,B.
Increasing the number of binding sites $N$ while keeping $M$ fixed can increase the Hill coefficient.
For fixed $N$, increasing $M$ changes the Hill coefficient in a nonmonotonic way:
there is an optimal value of $M$ that maximizes the Hill coefficient and therefore nonlinearity.
The effective inhibition threshold $K$ changes in a simpler form, increasing both with $N$ and $M$.
In conclusion, multiple binding sites can effectively increase the nonlinearity of the feedback via the regulatory function.

As discussed above, nonlinearity is an essential ingredient in a theory of biochemical oscillations~\cite{novak08}.
We therefore ask how multiple binding sites in the promoter affect a biochemical oscillator. 
We use the regulatory function Eq.~(\ref{eq.fnm}) in a generic model for genetic oscillations.
We consider a gene that encodes a protein that forms dimers, and these dimers can bind to multiple binding sites at the promoter to inhibit transcription, Fig.~\ref{fig.loop}A.
We introduce an explicit delay $\tau$ to account for transcription, translation, splicing, and other processes involved in the assembly of the gene product and its dimerization.
We assume that dimerization is a fast reaction, with a separation of timescales from other processes.
Therefore, at any time the dimer concentration can be approximated by that of the monomers squared.
The dynamics of the product concentration $x(t)$ is given by

\begin{equation}
\frac{dx}{dt}= b P \, \frac{  \sum\limits^{M}_{i=0}  {N \choose i} \left(  \frac{x(t-\tau)}{x_0} \right)^{2i} }{\left(1+   \left(\frac{x(t-\tau)}{x_0} \right)^2  \right)^{N}} - c \, x(t), 
\label{eq.oscillator}
\end{equation}
where $b$ is the basal production rate, $c$ is the decay rate of products, $P$ relates to the number of gene copies, 
$\tau$ is the total delay for product assembly, and $x_0$ is the product concentration that halves the basal transcription rate.
The regulatory function Eq.~(\ref{eq.fnm}) is parametrized by $N$ and $M$.
This genetic oscillator is a reduction of the models proposed by Lewis~\cite{lewis03}, and other authors~\cite{monk03,jensen03}.
It describes the protein concentration, but does not include the mRNA; the duration of transcription and translation are both included in the delay $\tau$.
Furthermore, it does not describe effects present in theories that include more than one regulator~\cite{cinquin07,schroter12,ay13}, 
but here it is enough to illustrate the effects of multiple binding sites in a simpler context.

We integrate Eq.~(\ref{eq.oscillator}) numerically and evaluate the resulting dynamics
by calculating the amplitude and period of oscillations. 
In all numerical simulations, we use the function \texttt{dde23} from MATLAB~\cite{shampine01}.
Scanning the values of $N$ and $M$, we determine whether the system oscillates in steady state:
when the difference in the maxima over the last ten cycles falls below 0.01, the simulation is stopped and we record the output.

The amplitude of oscillations grows with the number of binding sites $N$, Fig.~\ref{fig.step_fit}C.
The range where the system oscillates grows with the number of binding sites $N$.
The amplitude is nonmonotonic in $M$:
for a fixed number of binding sites $N$, there is an optimal value for $M$ that maximizes the amplitude of oscillations, Fig.\ref{fig.step_fit}C.
The period of oscillations grows with $N$ and decreases with $M$.
These results show how the change in nonlinearity observed in the regulatory functions
is reflected in the oscillations as the number of binding sites change.

\section{Gradual inhibition}
There is strong evidence for the products of some cyclic genes of the zebrafish segmentation clock 
binding their own promoters and acting as transcriptional repressors~\cite{oates02, giudicelli07, trofka12, brend09, schroter12, oates12}.
However, we do not have detailed knowledge of how these transcriptional repressors affect transcription rates when bound to the promoters of cyclic genes.
There is evidence from transcriptional analysis of the \emph{Hes1} gene in mouse indicating that inhibition is gradual~\cite{takebayashi94}.
While the wildtype \emph{Hes1} promoter containing all three N Box elements that are bound by Hes1 proteins showed a 30-fold inhibition of transcription in the presence of Hes1, 
mutations in one, two, and three of the N Box elements showed impaired inhibition with 14-, 7-, and 2-fold inhibition, respectively~\cite{takebayashi94}.

\begin{figure}[t]
\begin{center}
\includegraphics[width=\columnwidth]{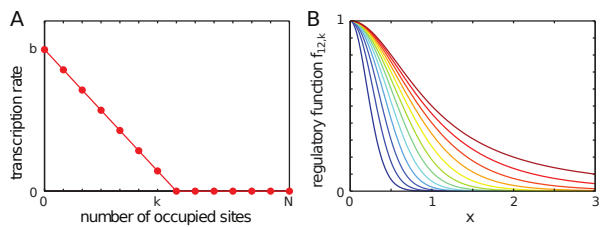}
\caption{Gradual inhibition. Binding of multiple transcriptional repressors gradually inhibits transcription. 
(A) Transcription rate as a function of the number of bound transcriptional repressors.
Transcription occurs at the basal rate $b$ in the absence of bound repressors, and decreases linearly to zero for more than $k$ bound repressors.
(B) Normalized regulatory function Eq.~(\ref{eq.fnk}) as a function of repressor concentration $x$, with $N=12$ and $k = 0,\ldots,11$ (dark blue to dark red).
}
\label{fig.gradual}
\end{center}
\end{figure}

\begin{figure}[t]
\begin{center}
\includegraphics[width=\columnwidth]{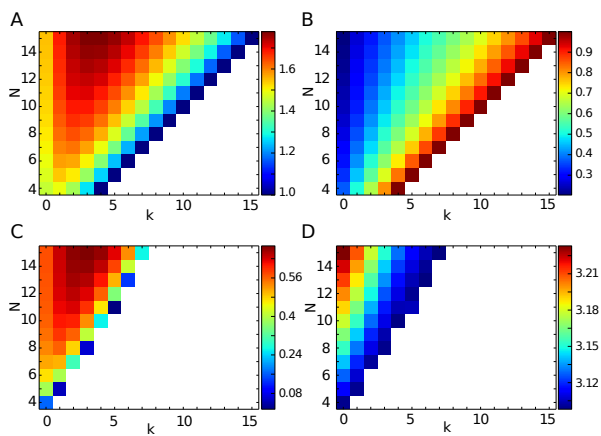}
\caption{Multiple binding sites can increase nonlinearity and enhance oscillations in the presence of gradual inhibition.
(A) and (B) Effective Hill parameters for the regulatory function $f_{N,k}$, Eq.~(\ref{eq.fnk}), with $b\,P=1$ and $x_0=1$.
(A) Effective Hill coefficient $h$.
(B) Effective inhibition threshold $K$.
(C) and (D) Oscillations obtained using regulatory functions Eq.~(\ref{eq.fnk}) with $b\,P= 2$, $\tau=1$, $c=1$ and $x_0=1$.
(C) Amplitude of oscillations.
(D) Period of oscillations.
In C and D the white region represents the nonoscillatory regime, in which the system settles to a fixed point.
Color bar labels indicate values in each panel.}
\label{fig.gradual_fit}
\end{center}
\end{figure}

Motivated by these results, we consider here a scenario where additional bound repressors gradually reduce transcription rate until it drops to zero, Fig.~\ref{fig.gradual}A.
For the sake of simplicity, we assume that transcription rate drops linearly as a function of bound repressors, from a basal rate $b$, to zero for $k+1$ occupied sites 

\begin{equation} \label{eq.gradual_inhibition}
a_i = \left\{	\begin{array}{ll}
		b \, \left( 1 - i / (k+1) \right)	& \textrm{if } i \leq k+1	\\
		0					& \textrm{if } i > k+1		\, ,
		\end{array} \right.
\end{equation}
see Fig.~\ref{fig.gradual}A.
Using this gradual inhibition in Eq.~(\ref{eq.fx}) together with Eq.~(\ref{eq.pipo}) and (\ref{eq.pop}), we obtain regulatory functions $f_{N,k}(x)$

\begin{align}		\label{eq.fnk}
f_{N,k}(x) &= b P \left( 1 + \frac{x}{x_0} \right)^{-N} \notag\\
&\times \sum_{i=0}^{k+1}{ \left( 1 - \frac{i}{k+1} \right) {N \choose i} \left( \frac{x}{x_0} \right)^{i}  } \, ,
\end{align}
see Fig.~\ref{fig.gradual}B. 
As in the previous case, it is clear that the effective inhibition threshold shifts to the right as $k$ increases, 
but it is not so clear if the steepness of the regulatory function changes and, if so, how.
Performing fits to Hill functions, we find that the nonmonotonic behavior of the effective Hill exponent $h$ is observed again as $k$ increases, Fig.~\ref{fig.gradual_fit}A,B.
Oscillations are similarly affected by noncooperative binding with gradual inhibition, Fig.~\ref{fig.gradual_fit}C,D.
These results show that the prediction of an optimal value for the number of bound repressors that fully inhibits the promoter is robust with respect to the details of how multiple bound repressors reduce the transcription rate.

\section{Discussion}
We studied the effects of multiple noncooperative binding sites in the promoter of a genetic oscillator.
We evaluated the behavior of a promoter with multiple binding sites when binding of transcriptional repressors is noncooperative, Fig.~\ref{fig.promoter}. 
We considered two different hypotheses for how bound transcriptional repressors affect transcription rates, Figs.~\ref{fig.step} and \ref{fig.gradual}.
In both cases, we calculated how the number of binding sites and the number of bound repressors required to produce full inhibition affect the nonlinearity of regulatory functions.
We showed that there is an optimal value of the number of repressors needed to fully inhibit transcription that maximizes nonlinearity, Figs.~\ref{fig.step_fit}AB and \ref{fig.gradual_fit}AB.
This increased nonlinearity is reflected in the behavior of a genetic oscillator controlled by such regulatory functions, Figs.~\ref{fig.step_fit}CD and \ref{fig.gradual_fit}CD.

Cooperative binding is a well known means to increase the nonlinearity of a biological dynamical system~\cite{ferrell09,qian12}. 
Here we show that the nonlinearity of a regulatory function can be increased by multiple binding sites, even if binding is noncooperative.
This idea may have application in other biological control systems as well.
Using the same formulation as we did here, one can also describe nonlinearity in the transcriptional activation of gene expression, 
thereby creating effective on-switches in developmental and physiological regulatory networks. 
A similar effect has been reported in a theoretical study of enzymes with multiple phosphorylation sites~\cite{wang10,ryerson13,enciso13}.
It was found that nonessential phosphorylation sites give rise to an increase in effective Hill coefficients, enhancing ultrasensitivity in signal transduction~\cite{enciso13}.

Previous work on the segmentation clock addressed the case of three binding sites~\cite{zeiser06}, motivated by experimental observations of the \emph{Hes7} mouse promoter~\cite{bessho01}.
The authors assumed that a single bound dimer inhibits transcription completely, corresponding to the particular case $M=1$ of the theory developed here.
They considered both noncooperative and cooperative binding, and showed that cooperativity increases the effective Hill coefficient as expected.
In a follow-up~\cite{zeiser07}, the authors addressed the case of \emph{Hes1} regulation based on the report of four binding sites in the \emph{Hes1} mouse promoter~\cite{takebayashi94}.
Again assuming that a single bound dimer inhibits transcription completely, they used the data from the transcriptional analysis of the \emph{Hes1} gene 
to estimate an effective Hill coefficient for the Mouse Hes1 oscillator, obtaining an upper bound of about $3$.
More recently, the effect of two binding sites as compared to a single binding site was discussed together with differential decay of the monomers~\cite{campanelli10}.

Our theory predicts how changing the number of binding sites $N$, and the number $M$ of bound repressors that produce full inhibition affects a single cell oscillator.
Although an experiment that changes the value of $M$ may currently be challenging in a cellular system, the number of binding sites $N$ is more amenable to experimental manipulation.
For example, binding sites could be mutated~\cite{takebayashi94}, or deleted from the promoter, 
or they could be interfered with using genome editing strategies such as TALEN~\cite{badell12} or CRISPR~\cite{pennisi13} to alter or delete specific binding sites.
To assess the effects of these perturbations in experiments may also pose some challenges.
Dropping the number of binding sites from $N=12$ to $N=6$ introduces a period change of about 2.5\%, while amplitude halves over the same range, Fig.~\ref{fig.step_fit}C,D.
Experiments will require at least such precision to reliably detect changes.

Our results suggest a possible evolutionary mechanism to increase nonlinearity in gene regulatory systems.
In this mechanism, point mutations in the promoter that increase the number of binding sites for transcription factors may increase the steepness of regulatory functions.
If the resulting steeper regulation performs some function better, such mutations would have a good chance to be conserved by natural selection.
In the case of the segmentation clock, the amplitude of oscillations could increase with the number of binding sites, possibly reducing the signal to noise ratio.
Furthermore, the range for oscillations would be wider, making the oscillatory regime less sensitive to slow extrinsic fluctuations of parameter values.
Remarkably, it may happen that after an increase in $N$, an increase in $M$ also raises nonlinearity, see Fig.~\ref{fig.step_fit}A.
This means that weaker repressors would result in better oscillations.
This evolutionary mechanism would provide a simple way to gradually increase the nonlinearity of a feedback or other regulatory function.

The theory for the zebrafish regulatory function could be refined using the experimentally measured relative affinities of the binding sites at the \emph{her1} and \emph{her7} promoters~\cite{schroter12}.
Apart from the effects reported here, the number of binding sites may have additional roles.
For example, it could serve as a buffer for fluctuations in gene expression~\cite{barkai00,elowitz02,raser05,munsky12,tsimring14,morelli07}, augmenting the precision of genetic oscillations.
This will be the topic of future work.

\begin{acknowledgements}
We thank Sa\'ul Ares, Luciana Bruno, Ariel Chernomoretz and Hernan Grecco for helpful comments on an early version of this work, and James Ferrell for calling our attention to Ref.~\cite{wang10}. Thanks to E. Aikau for inspiration. LGM acknowledges support from MINCyT Argentina (PICT 2012 1952). A.C.O. and D.S. were supported by the Medical Research Council UK (MC\_UP\_1202/3) and the Wellcome Trust (WT098025MA).
\end{acknowledgements}

\begin{figure}[tb]
\begin{center}
\includegraphics[width=\columnwidth]{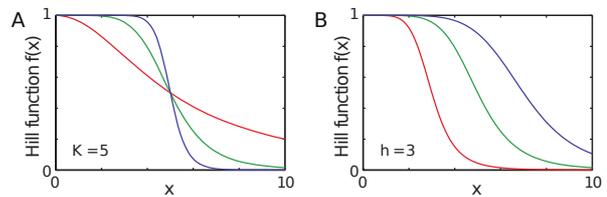}
\caption{Hill functions are characterized by two parameters, the Hill coefficient $h$ and the inhibition threshold $K$.
(A) Hill functions with $K=5$ and $h=1$ (red), $h=3$ (green) and $h=7$ (blue).
(B) Hill functions with $h=3$ and $K=3$ (red), $K=5$ (green) and $K=7$ (blue).
}
\label{fig.hill}
\end{center}
\end{figure}

\section*{Appendix: Effective Hill functions}
\begin{figure}[tb]
\begin{center}
\includegraphics[width=\columnwidth]{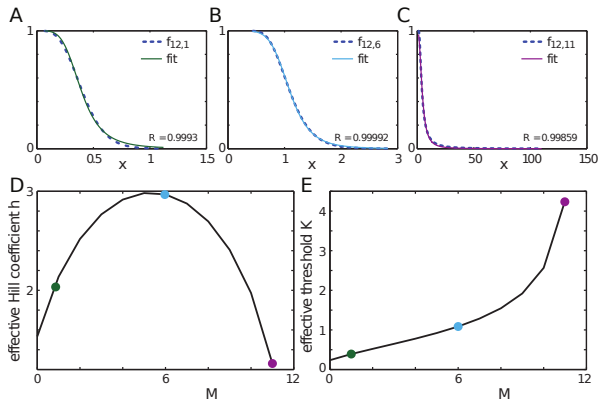}
\caption{Effective Hill functions Eq.~(\ref{eq.hill}) can fit regulatory functions of the multiple binding site regulatory function, 
Eq.~(\ref{eq.fnm}) for $N=12$, with $b\,P=1$ and $x_0=1$.
(A-C) Examples of fits of Hill functions (solid lines) to regulatory functions (dashed lines) for 
(A) $M=1$ (green), (B) $M=6$ (light blue) and (C) $M=11$ (purple).
(D) Effective Hill coefficient $h$ as a function of $M$.
(E) Effective inhibition threshold $K$ as a function of $M$.
Dots in D and E correspond to fits from panels A, B and C.
}
\label{fig.hill_fit}
\end{center}
\end{figure}

Hill functions are often used to describe the nonlinearities present in gene regulatory networks~\cite{alon06}.
Hill functions are sigmoidal step functions defined by

\begin{equation}
f_{H}(x) = \frac{1}{1 + \left( {x / K} \right)^{2 h}},
\label{eq.hill}
\end{equation}  
where the steepnes of the step is characterized by the exponent $2 h$, 
and the inhibition threshold $K$ is the concentration of repressor that halves the production rate, here scaled to unity, Fig.~\ref{fig.hill}.
Here we include an explicit factor $2$ in the exponent to account for the dimerization of the transcriptional repressors.
One advantage of Hill functions is that they are very simply parametrized.
In the main text, we fit Hill functions to the more complex regulatory functions Eqs.~(\ref{eq.fnm}) and~(\ref{eq.fnk}).
Some fits of Eq.~(\ref{eq.fnm}) in the case $N=12$ are displayed in Fig.~\ref{fig.hill_fit} for illustration.

\begin{thebibliography}{10}

\bibitem{goldbeter}
A~Goldbeter,
\newblock {\it Biochemical Oscillations and Cellular Rhythms: The Molecular
  Bases of Periodic and Chaotic Behaviour},
\newblock Cambridge University Press, Cambridge (1997).

\bibitem{mackey}
L~Glass, M C Mackey,
\newblock {\it From Clocks to Chaos: The Rhythms of Life},
\newblock Princeton University Press, Princeton (1988).

\bibitem{liu95}
Y~Liu, N F Tsinoremas, C H Johnson, N V Lebedeva, S S G M Ishiura, T~Kondo,
\newblock {\it Circadian orchestration of gene expression in cyanobacteria},
\newblock Gene. Dev. {\bf 9}, 1469 (1995).

\bibitem{nagoshi04}
E~Nagoshi, C~Saini, C~Bauer, T~Laroche, F~Naef, U~Schibler,
\newblock {\it Circadian gene expression in individual fibroblasts:
  Cell-autonomous and self-sustained oscillators pass time to daughter cells},
\newblock Cell {\bf 119}, 693 (2004).

\bibitem{mihalcescu04}
I~Mihalcescu, W~Hsing, S~Leibler,
\newblock {\it Resilient circadian oscillator revealed in individual
  cyanobacteria},
\newblock Nature {\bf 430}, 81 (2004).

\bibitem{goldbeter12}
A~Goldbeter, C~G{\'e}rard, D~Gonze, J C Leloup, G~Dupont,
\newblock {\it Systems biology of cellular rhythms},
\newblock FEBS Lett. {\bf 586}, 2955 (2012).

\bibitem{palmeirim97}
I~Palmeirim, D~Henrique, D~Ish-Horowicz, O~Pourqui\'{e},
\newblock {\it Avian {\it hairy} gene expression identifies a molecular clock
  linked to vertebrate segmentation and somitogenesis},
\newblock Cell {\bf 91}, 639 (1997).

\bibitem{aulehla08}
A~Aulehla, W~Wiegraebe, V~Baubet, M B Wahl, C~Deng, M~Taketo, M~Lewandoski,
  O~Pourqui\'{e},
\newblock {\it A $\beta$-catenin gradient links the clock and wavefront systems
  in mouse embryo segmentation},
\newblock Nat. Cell Biol. {\bf 10}, 186 (2008).

\bibitem{masamizu06}
Y~Masamizu, T~Ohtsuka, Y~Takashima, H~Nagahara, Y~Takenaka, K~Yoshikawa,
  H~Okamura, R~Kageyama,
\newblock {\it Real-time imaging of the segmentation clock: Revelation of
  unstable oscillators in the individual presomitic mesoderm cells},
\newblock Proc. Natl. Acad. Sci. USA {\bf 103}, 1313 (2006).

\bibitem{krol11}
A J Krol, D~Roellig, M L Dequ{\'e}ante, O~Tassy, E~Glynn, G~Hattem,
  A~Mushegian, A C Oates, O~Pourqui{\'{e}},
\newblock {\it Evolutionary plasticity of segmentation clock networks},
\newblock Development {\bf 138}, 2783 (2011).

\bibitem{shimojo08}
H~Shimojo, T~Ohtsuka, R~Kageyama,
\newblock {\it Oscillations in notch signaling regulate maintenance of neural
  progenitors},
\newblock Neuron {\bf 58}, 52 (2008).

\bibitem{geva06}
N~Geva-Zatorsky, N~Rosenfeld, S~Itzkovitz, R~Milo, A~Sigal, E~Dekel,
  T~Yarnitzky, Y~Liron, P~Polak, G~Lahav, U~Alon,
\newblock {\it Oscillations and variability in the p53 system},
\newblock Mol. Syst. Biol. {\bf 2}, 2006.0033 (2006).

\bibitem{elowitz00}
M B Elowitz, S~Leibler,
\newblock {\it A synthetic oscillatory network of transcriptional regulators},
\newblock Nature {\bf 403}, 335 (2000).

\bibitem{stricker08}
J~Stricker, S~Cookson, M R Bennett, W H Mather, L S Tsimring, J~Hasty,
\newblock {\it A fast, robust and tunable synthetic gene oscillator},
\newblock Nature {\bf 456}, 516 (2008).

\bibitem{tyson.in.keizer}
J J Tyson,
\newblock {\it Computational cell biology}, Chap. 9, Pag. 230,
\newblock Springer, Berlin (2002).

\bibitem{alberts}
B~Alberts, A~Johnson, J~Lewis, M~Raff, K~Roberts, P~Walter,
\newblock {\it Molecular biology of the cell}, 4$^{th}$ Ed.,
\newblock Garland Science, New York (2002).

\bibitem{alon06}
U~Alon,
\newblock {\it An introduction to systems biology: Design principles of
  biological circuits},
\newblock Chapman \& Hall/CRC Press, Boca Raton, Florida (2006).

\bibitem{novak08}
B~Nov{\'{a}}k, J J Tyson,
\newblock {\it Design principles of biochemical oscillators},
\newblock Nat. Rev. Mol. Cell Biol. {\bf 9}, 981 (2008).

\bibitem{lewis03}
J~Lewis,
\newblock {\it Autoinhibition with transcriptional delay: A simple mechanism
  for the zebrafish somitogenesis oscillator},
\newblock Curr. Biol. {\bf 13}, 1398 (2003).

\bibitem{morelli07}
L G Morelli, F~J{\"{u}}licher,
\newblock {\it Precision of genetic oscillators and clocks},
\newblock Phys. Rev. Lett. {\bf 98}, 228101 (2007).

\bibitem{ferrell09}
J E Ferrell,
\newblock {\it Q\&A: Cooperativity},
\newblock J. Biol. {\bf 8}, 53 (2009).

\bibitem{pourquie11}
O~Pourqui{\'e},
\newblock {\it Vertebrate segmentation: From cyclic gene networks to
  scoliosis},
\newblock Cell {\bf 145}, 650 (2011).

\bibitem{oates12}
A C Oates, L G Morelli, S~Ares,
\newblock {\it Patterning embryos with oscillations: Structure, function and
  dynamics of the vertebrate segmentation clock},
\newblock Development {\bf 139}, 625 (2012).

\bibitem{saga12b}
Y~Saga,
\newblock {\it The synchrony and cyclicity of developmental events},
\newblock Cold Spring Harb. Perspect. Biol. {\bf 4}, a008201 (2012).

\bibitem{roellig11}
D~Roellig, L G Morelli, S~Ares, F~J{\"{u}}licher, A C Oates,
\newblock {\it Snapshot: The segmentation clock},
\newblock Cell {\bf 145}, 800 (2011).

\bibitem{saga12a}
Y~Saga,
\newblock {\it The mechanism of somite formation in mice},
\newblock Curr. Opin. Genet. Dev. {\bf 22}, 331 (2012).

\bibitem{oates02}
A C Oates, R K Ho,
\newblock {\it {\it Hairy/E(spl)-related} ({{\it Her}}) genes are central
  components of the segmentation oscillator and display redundancy with the
  {Delta/Notch} signaling pathway in the formation of anterior segmental
  boundaries in the zebrafish},
\newblock Development {\bf 129}, 2929 (2002).

\bibitem{hirata02}
H~Hirata, S~Yoshiura, T~Ohtsuka, Y~Bessho, T~Harada, K~Yoshikawa, R~Kageyama,
\newblock {\it Oscillatory expression of the {bHLH factor Hes1} regulated by a
  negative feedback loop},
\newblock Science {\bf 298}, 840 (2002).

\bibitem{holley02}
S A Holley, D~J{\"{u}}lich, G J Rauch, R~Geisler, C~N{\"{u}}sslein-Volhard,
\newblock {\it {\it her1} and the {\it notch} pathway function within the
  oscillator mechanism that regulates zebrafish somitogenesis},
\newblock Development {\bf 129}, 1175 (2002).

\bibitem{takebayashi94}
K~Takebayashi, Y~Sasai, Y~Sakai, T~Watanabe, S~Nakanishi, R~Kageyama,
\newblock {\it Structure, chromosomal locus, and promoter analysis of the gene
  encoding the mouse helix-loop-helix factor hes-1. negative autoregulation
  through the multiple n box elements},
\newblock J. Biol. Chem. {\bf 269}, 5150 (1994).

\bibitem{bessho01}
Y~Bessho, G~Miyoshi, R~Sakata, R~Kageyama,
\newblock {\it Hes7: A bhlh-type repressor gene regulated by notch and
  expressed in the presomitic mesoderm},
\newblock Genes Cells {\bf 6}, 175 (2001).

\bibitem{schroter12}
C~Schr{\"{o}}ter, S~Ares, L G Morelli, A~Isakova, K~Hens, D~Soroldoni,
  M~Gajewski, F~J{\"{u}}licher, S J Maerkl, B~Deplancke, A C Oates,
\newblock {\it Topology and dynamics of the zebrafish segmentation clock core
  circuit},
\newblock PLoS Biol. {\bf 10}, e1001364 (2012).

\bibitem{trofka12}
A~Trofka, J~Schwendinger-Schreck, T~Brend, W~Pontius, T~Emonet, S A Holley,
\newblock {\it The {Her7} node modulates the network topology of the zebrafish
  segmentation clock via sequestration of the {Hes6} hub},
\newblock Development {\bf 139}, 940 (2012).

\bibitem{hanisch13}
A~Hanisch, M V Holder, S~Choorapoikayil, M~Gajewski, E M \"Ozbudak, J~Lewis,
\newblock {\it The elongation rate of {RNA} polymerase ii in zebrafish and its
  significance in the somite segmentation clock},
\newblock Development {\bf 140}, 444 (2013).

\bibitem{giudicelli07}
F~Giudicelli, E M {\"O}zbudak, G J Wright, J~Lewis,
\newblock {\it Setting the tempo in development: An investigation of the
  zebrafish somite clock mechanism},
\newblock PLoS Biol. {\bf 5}, 1309 (2007).

\bibitem{ozbudak08}
E M {\"O}zbudak, J~Lewis,
\newblock {\it Notch signalling synchronizes the zebrafish segmentation clock
  but is not needed to create somite boundaries},
\newblock PLoS Genet. {\bf 4(2)}, e15 (2008).

\bibitem{harima13}
Y~Harima, Y~Takashima, Y~Ueda, T~Ohtsuka, R~Kageyama,
\newblock {\it Accelerating the tempo of the segmentation clock by reducing the
  number of introns in the hes7 gene},
\newblock Cell Rep. {\bf 3}, 1 (2013).

\bibitem{keener}
J~Keener, J~Sneyd,
\newblock {\it Mathematical physiology I: Cellular physiology}, 2$^{nd}$ Ed.
\newblock Springer, Berlin (2008).

\bibitem{qian12}
H~Qian,
\newblock {\it Cooperativity in cellular biochemical processes: Noise-enhanced
  sensitivity, fluctuating enzyme, bistability with nonlinear feedback, and
  other mechanisms for sigmoidal responses},
\newblock Ann. Rev. Biophys. {\bf 41}, 179 (2012).

\bibitem{zeiser06}
S~Zeiser, H V Liebscher, H~Tiedemann, I~Rubio-Aliaga, G K H Przemeck, M H
  de~Angelis, G~Winkler,
\newblock {\it Number of active transcription factor binding sites is essential
  for the hes7 oscillator},
\newblock Theor. Biol. Med. Model. {\bf 3}, 11 (2006).

\bibitem{gunawardena05}
J~Gunawardena,
\newblock {\it Multisite protein phosphorylation makes a good threshold but can
  be a poor switch},
\newblock P. Natl. Acad. Sci. USA {\bf 102}, 14617 (2005).

\bibitem{gotea10}
V~Gotea, A~Visel, J M Westlund, M A Nobrega, L A Pennacchio, I~Ovcharenko,
\newblock {\it Homotypic clusters of transcription factor binding sites are a
  key component of human promoters and enhancers},
\newblock Genome Res. {\bf 20}, 565 (2010).

\bibitem{burz98}
D S Burz, R~Rivera-Pomar, H~J{\"a}ckle, S D Hanes,
\newblock {\it Cooperative dna-binding by bicoid provides a mechanism for
  threshold-dependent gene activation in the drosophila embryo},
\newblock EMBO J. {\bf 17}, 5998 (1998).

\bibitem{brend09}
T~Brend, S A Holley,
\newblock {\it Expression of the oscillating gene her1 is directly regulated by
  hairy/enhancer of split, t-box, and suppressor of hairless proteins in the
  zebrafish segmentation clock},
\newblock Dev. Dynam. {\bf 238}, 2745 (2009).

\bibitem{soroldoniPC}
\newblock D~Soroldoni, personal communication (2014).

\bibitem{bintu05a}
L~Bintu, N E Buchler, H G Garcia, U~Gerland, T~Hwa, J~Kondev, R~Phillips,
\newblock {\it {Transcriptional regulation by the numbers: Models}},
\newblock Curr. Opin. Genet. Dev. {\bf 15}, 116 (2005).

\bibitem{garcia10}
H G Garcia, A~Sanchez, T~Kuhlman, J~Kondev, R~Phillips,
\newblock {\it Transcription by the numbers redux: experiments and calculations
  that surprise},
\newblock Trends Cell Biol. {\bf 20}, 723 (2010).

\bibitem{monk03}
N A M Monk,
\newblock {\it Oscillatory expression of {Hes1, p53, and NF}-$\kappa${B} driven
  by transcriptional time delays},
\newblock Curr. Biol. {\bf 13}, 1409 (2003).

\bibitem{jensen03}
M H Jensen, K~Sneppen, G~Tiana,
\newblock {\it Sustained oscillations and time delays in gene expression of
  protein {Hes1}},
\newblock FEBS Lett. {\bf 541}, 176 (2003).

\bibitem{cinquin07}
O~Cinquin,
\newblock {\it Repressor dimerization in the zebrafish somitogenesis clock},
\newblock PLoS Comp. Biol. {\bf 3}, e32 (2007).

\bibitem{ay13}
A~Ay, S~Knierer, A~Sperlea, J~Holland, E M {\"O}zbudak,
\newblock {\it Short-lived her proteins drive robust synchronized oscillations
  in the zebrafish segmentation clock},
\newblock Development {\bf 140}, 3244 (2013).

\bibitem{shampine01}
L F Shampine, S~Thompson,
\newblock {\it Solving ddes in Matlab},
\newblock Appl. Numer. Math. {\bf 37}, 441 (2001).

\bibitem{wang10}
L~Wang, Q~Nie, G~Enciso,
\newblock {\it Nonessential sites improve phosphorylation switch},
\newblock Biophys. J. {\bf 99}, L41 (2010).

\bibitem{ryerson13}
S~Ryerson, G~Enciso,
\newblock {\it Ultrasensitivity in independent multisite systems},
\newblock J. Math. Biol. {\bf 69}, 977 (2014).

\bibitem{enciso13}
G~Enciso,
\newblock {\it Nonautonomous and random dynamical systems in life sciences}
\newblock Lecture Notes in Mathematics (Mathematical Biosciences
  Subseries) No 2102, Springer Verlag, Berlin (2013).

\bibitem{zeiser07}
S~Zeiser, J~M{\"u}ller, V~Liebscher,
\newblock {\it Modeling the hes1 oscillator},
\newblock J. Comput. Biol. {\bf 14}, 984 (2007).

\bibitem{campanelli10}
M~Campanelli, T~Gedeon,
\newblock {\it Somitogenesis clock-wave initiation requires differential decay
  and multiple binding sites for clock protein},
\newblock PLoS Comp. Biol. {\bf 6}, e1000728 (2010).

\bibitem{badell12}
V M Bedell, Y~Wang, J M Campbell, T L Poshusta, C G Starker, R G K II, W~Tan, S
  G Penheiter, A C Ma, A Y H Leung, S C Fahrenkrug, D F Carlson, D F Voytas, K
  J Clark, J J Essner, S C Ekker,
\newblock {\it In vivo genome editing using a high-efficiency talen system},
\newblock Nature {\bf 491}, 114 (2012).

\bibitem{pennisi13}
E~Pennisi,
\newblock {\it The crispr craze},
\newblock Science {\bf 341}, 833 (2013).

\bibitem{barkai00}
N~Barkai, S~Leibler,
\newblock {\it Biological rhythms: Circadian clocks limited by noise},
\newblock Nature {\bf 403}, 267 (2000).

\raggedbottom
\pagebreak

\bibitem{elowitz02}
M B Elowitz, A J Levine, E D Siggia, P S Swain,
\newblock {\it Stochastic gene expression in a single cell},
\newblock Science {\bf 297}, 1183 (2002).

\bibitem{raser05}
J~Raser, E~O'Shea,
\newblock {\it Noise in gene expression: Origins, consequences, and control},
\newblock Science {\bf 309}, 2010 (2005).

\bibitem{munsky12}
B~Munsky, G~Neuert, A V Oudenaarden,
\newblock {\it Using gene expression noise to understand gene regulation},
\newblock Science {\bf 336}, 183 (2012).

\bibitem{tsimring14}
L S Tsimring,
\newblock {\it Noise in biology},
\newblock Rep. Prog. Phys. {\bf 77}, 026601 (2014).

\end{thebibliography}

\end{document}